\documentclass[journal]{IEEEtran}
\ifCLASSINFOpdf
   \usepackage[pdftex]{graphicx}
\else
\fi
\usepackage{amsmath}
\usepackage{amsfonts}
\usepackage[justification=centering]{caption}
\usepackage{url}

\begin{document}
%
\title{$hf_0$:  A hybrid pitch extraction method for multimodal voice}
%
%
%

\author{Pradeep~Rengaswamy, Gurunath~Reddy M
        and~Krothapalli~Sreenivasa~Rao}
\vspace{-2cm}

\maketitle

\begin{abstract}

Pitch or fundamental frequency ($f_0$) extraction is a fundamental problem studied extensively for its potential applications in speech and clinical applications. In literature, explicit mode specific (modal speech or singing voice or emotional/ expressive speech or noisy speech) signal processing and deep learning $f_0$ extraction methods that exploit the quasi periodic nature of the signal in time, harmonic property in spectral or combined form to extract the pitch is developed. Hence, there is no single unified method which can reliably extract the pitch from various modes of the acoustic signal. In this work, we propose a hybrid $f_0$ extraction method which seamlessly extracts the pitch across modes of speech production with very high accuracy required for many applications. The proposed hybrid model exploits the advantages of deep learning and signal processing methods to minimize the pitch detection error and adopts to various modes of acoustic signal. Specifically, we propose an ordinal regression convolutional neural networks to map the periodicity rich input representation to obtain the nominal pitch classes which drastically reduces the number of classes required for pitch detection unlike other deep learning approaches. Further, the accurate $f_0$ is estimated from the nominal pitch class labels by filtering and autocorrelation. We show that the proposed method generalizes to the unseen modes of voice production and various noises for large scale datasets. Also, the proposed hybrid model significantly reduces the learning parameters required to train the deep model compared to other methods. Furthermore, the evaluation measures showed that the proposed method is significantly better than the state-of-the-art signal processing and deep learning approaches.

\end{abstract}


%
\IEEEpeerreviewmaketitle

\begin{figure*}[t]
\centering
\includegraphics[width=\textwidth]{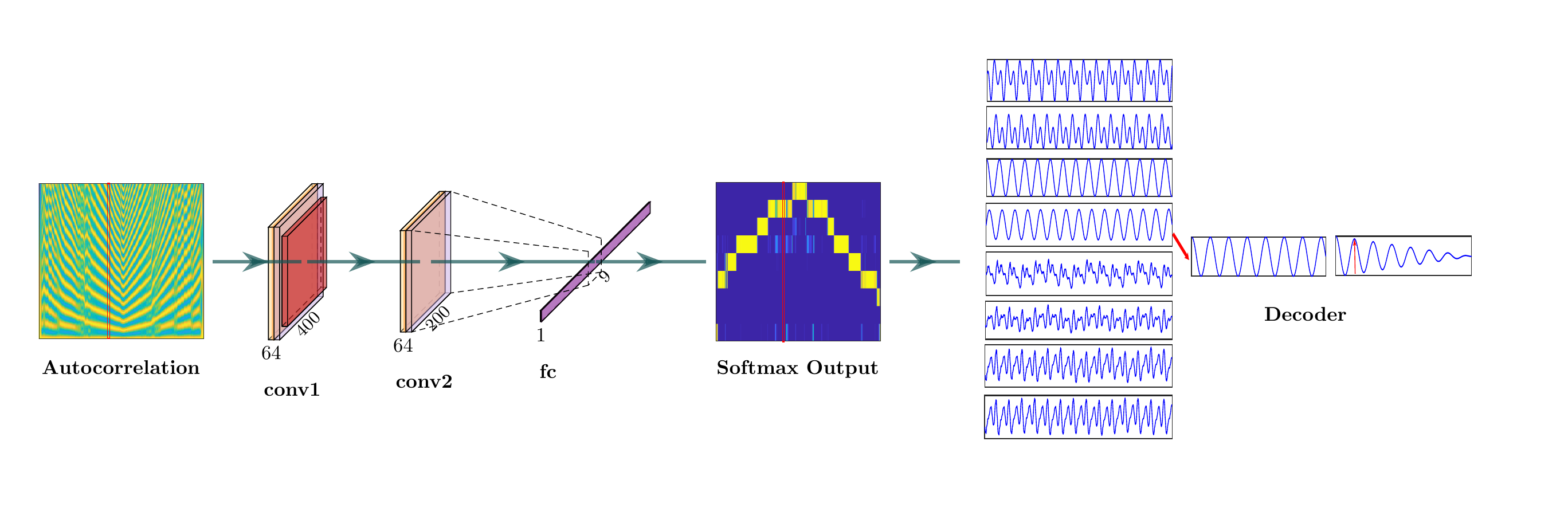}
\vspace{-1.5cm}
\caption{Overview of the proposed $hf_0$ pitch extraction method.}
\label{fig:overview}
\vspace{-0.2cm}

\end{figure*}

\section{Introduction}

Pitch is a perceptual quantity defined based on the auditory feedback of the human listener. The pulmonary and laryngeal configuration of the vocal tract system determines the pitch of the speaker or singer. We can quantify pitch as the fundamental frequency ($f_0$) based on the periodicity of the acoustic signal. We can find several methods which adopts $f_0$ extraction methods developed for monophonic speech to other modes by introducing mode specific information say for example, pYIN~\cite{mauch2014pyin} adopts the YIN~\cite{de2002yin} algorithm developed partially for monophonic speech to singing voice by explicitly modeling the distributions of singing voice parameters. Also, we can find that $f_0$ extraction method developed for one mode may not generalize to the other modes of acoustic signal. For example, $f_0$ extraction method developed for monophonic speech do not generalize for singing voice and vice-versa. This is due to significant variability of acoustic properties of different modes of voice production: speech, singing voice, emotional/expressive speech, and noisy speech. Therefore, we find no single unified approach which can reliably extract the pitch from various modes of acoustic signal. Hence, in this paper, we propose a hybrid $f_0$ extraction method which seamlessly extracts the pitch from various modes of human acoustic voice production with very high accuracy required for many potential applications: vocal pedagogy~\cite{sataloff2006vocal}, singer identification~\cite{nwe2007exploring}, synthesis of songs and speech~\cite{macon1997singing,saino2006hmm,saitou2005development,kenmochi2007vocaloid}, melody extraction~\cite{salamon2012melody},  expressive/emotional speech synthesis/conversion~\cite{schroder2001emotional}, paralinguistic voice analysis~\cite{ishi2008automatic}, and audio event detection~\cite{kumar2012audio}. In medical therapy, the $f_0$ extraction helps in emotion recognition~\cite{quam2012development} and clinical diagnosis of vocal disorders~\cite{brockmann2017acoustic,lombardo2016analysis} such as Dysphonia~\cite{pylypowich2016differentiating,lv2017serious} and many more. Furthermore, in the treatment of neuro-degenerative disorders such as Autism~\cite{hardy2013rhythm}, Alzheimer~\cite{maclean2017unforgettable} and Dementia~\cite{vink2003music} by detecting suitable fundamental frequency signal for auditory cueing~\cite{schaefer2014auditory}.

 
In this paper, we propose a novel hybrid $f_0$ extraction method which exploits the advantages of both traditional signal processing and deep learning approaches. The proposed hybrid $f_0$ extractor hereafter $hf_0$ is designed to extract the $f_0$ robustly and seamlessly from most common modes of voice production such as speech, monophonic singing voice, emotional/expressive speech, paralinguistic (laughter) and creaky voice, and noisy speech. In this letters, we propose a unified $f_0$ extraction method for multiple modalities of voice production, which is unexplored before to the best of our knowledge. The main contributions of this letters includes i) a hybrid $f_0$  extraction method $hf_0$ which exploits the advantages of deep learning and signal processing methods to minimize the pitch detection error and adopts to various modes of acoustic voice production. ii) We propose an ordinal regression convolutional neural networks to map the periodicity rich input representation into decodable nominal pitch classes which drastically reduces the number of classes required for pitch detection unlike other deep learning methods. iii) Pitch frequency band is obtained from the nominal pitch class labels to obtain the narrow band time domain signal to obtain the pitch. 
iv) We show that the proposed method generalizes to the unseen modes of voice production and various noises for large scale datasets. v) The proposed hybrid model significantly reduces the learning parameters required to train the deep model compare to other methods thus achieving significant model compression for mobile applications. vi) The proposed method is made open-source for  reproducible research at \textit{\url{https://github.com/Pradeepiit/hf0}}. The qualitative and quantitative evaluation measures showed that the proposed method is significantly better than the state-of-the-art signal processing and deep learning approaches for various modes of voice production.

\section{$hf_0$: Pitch Extraction Method}

Overview of the proposed $hf_0$ is shown in Fig.~\ref{fig:overview}. $hf_0$ consists of  i) an encoder which encodes the plausible pitch frequency band of the input signal by supervised learning model, and ii) a decoder which decodes the pitch unsupervisedly by signal processing approach. 

\subsection{Supervised Nominal Pitch Class Encoder}
\label{subsec:sup_pitch_freq_band_encode}
The existing supervised classification based pitch detection methods~\cite{liu2017novel, han2014neural, verma2016frequency} discretize the available pitch frequency range on the musically relevant logarithm scale to treat the pitch detection as a classification problem. Generally, the available pitch frequency range is quantized into 67 levels which covers the plausible pitch frequency range of  $60~-~400~Hz$ and an additional voicing label to classify each frame belonging to voiced or unvoiced. Similarly, to accommodate the wide pitch frequency range of the singing voice, the pitch frequency range is divided into 360 levels in CREPE~\cite{kim2018crepe} to achieve good pitch resolution. Thus, we can conclude that as the pitch frequency range increases, the number of class will also increases to preserve good pitch resolution. Hence, the classification model trained for speech is inadequate to extract the pitch from wide band singing voice where most of the class labels of  singing voice becomes redundant for speech as pitch frequency range of speech is relatively lower compared to singing voice results in very weak posterior probabilities assigned to each class label, results in multiple-misclassifications. Also, as the number of classes increases, the number of parameters required to train the model also increases to fit the very complex model and also we need a large amount of data to train such models because the extreme class labels belonging to very low and very high pitch values gets very less labels resulting in models bias towards the mid-frequency range. To alleviate this probable, we treat sub-bands of the plausible pitch range as class which reduces the number classes required significantly and the amount of data required to train the models. Hence, generalizes to multiple modes of voice production. Specifically, we divide the available pitch frequency range into sub-bands and treat these bands as classes and an additional label for voicing decision.   

\subsubsection{Input and Output Representation}
\label{subsubsec:input_output}

The input audio signal is sampled at $16~kHz$ sampling rate with analysis frame size of $50~ms$ and $80~\%$ overlap between successive frames. The auto-correlation provides phase-normalization by converting phase shifted time-domain signal into zero phase cosine signal modulating at fundamental frequency. 
Thus, the periodicity of the voiced frames are enhanced by computing the auto-correlation ($r_t(\tau)$) of the signal at lags of $\tau$.

\begin{equation}\label{eq1}
r_t(\tau) = \frac{1}{N}\sum_{j=t+1}^{j=t+N} x_t(j) x_t(j+\tau), 0\leq \tau < N
\end{equation}

\noindent where N is the length of the frame, $x_t$ is the $t^{th}$ frame of the audio signal. Each frame $r_t(\tau)$ is energy normalized $nr_{t} = $ $r_t(\tau)/r_t(0)$ to diminish the model bias towards the energy. 

The normalized feature vector $nr_{t}$ is fed to the CNN to compute the posterior probability of the pitch frequency band states of a frame $t$ i.e., $p(y_t/nr_{t})$ where $y_t$ represents the set of  frequency bands of frame at time $t$. $y_t$ will contain eight unique frequency band sates $\{s_1, s_2, ..., s_8\}$ along with an additional voicing state $v$ correspond to voiced or silence state. The states from $s_1$ to $s_8$ corresponds to different frequency bands ranging from $50~Hz~to~800~Hz$. Specifically, the pitch frequency range ($50~Hz - 800~Hz$) is divided into eight frequency bands such as $[50-75~Hz)$, $[75-100~Hz)$, $[100-150~Hz)$, $[150-200~Hz)$, $[200-300~Hz)$, $[300-400~Hz)$, $[400-600~Hz)$, and $[600-800~Hz)$ for each frame which corresponds to frequency band states ${s_1, ..., s_8}$ respectively. Along with voicing state, we treat eight states ${s_1, ..., s_8}$ as minimal class labels to the CNN model i.e., for each frame, we create a one-hot vector as targets to the CNN by assigning magnitude = 1 for state $s_i$ if the ground truth pitch falls in $s_i$ frequency band otherwise it is assigned with magnitude = 0. The model predicts the posterior probability $p(y_t = s_i/nr_t)$ = 1 if the ground truth pitch falls in the frequency band $s_i$.  

\subsubsection{CNN Model}
\label{subsubsec:cnn_model}

The proposed shallow CNN model to obtain nominal pitch band class labels is shown in Fig.~\ref{fig:overview}. The autocorrelation coefficients of two neighbouring frames (two predecessor, current and two successor frames) are fed as input to the two layer CNN network. The 2048 dimensional latent representation of the CNN model is then connected densely to the output layer with softmax activation which corresponds to 9-dimensional output vector. Further, from the 9-nominal class labels, the $f_0$ is decoded unsupervisedly. The convolutional layers apply 64 filters with small receptive fields of size 3x3, stride 1 and zero padding to retain the size of the input. The activations of the convolutional layers are passed through RELU non-linear activations succeeded by batch normalization~\cite{ioffe2015batch}. The max-pooling of size 2x2 and stride 2 is performed after batch normalization layer only for the first convolutional layer. A dropout  layer with 0.2 probability is used for both convolutional layers to reduce overfitting. The output of the second dropout layer is flattened and provided as input to a dense layer with nine neurons. Further, normalized probability distribution for the class labels is attained by softmax activation.  As stochastic gradient descent (SGD)~\cite{wilson2017marginal} provides better generalization than the adaptive optimizers, SGD is used as the optimizer with momentum of 0.9, learning rate 0.001 and mini-batch size of 64 samples.

\subsection{Unsupervised Pitch Decoder}
\label{subsec:unsup_pitch_dec}
We obtain the narrow band pass signal which contains the $f_0$  information from the time-domain acoustic signal from the predicted frequency band class labels. Further, the pitch in the decoded signal bands is obtained by double autocorrelation which alleviates the need for additional post-processing method such as Viterbi decoding used by most of the popular  $f_0$ extraction methods to improve the accuracy. 

\subsubsection{Pitch Frequency Band Decoding}
\label{subsubsec:pitch_freq_band_decod}

The time domain acoustic signal $x_t(n)$ is decomposed into narrow band signals with narrow band pass filter using predicted frequency band class labels. The filter $g_{b}(n)$ represents a band-pass filter with minimum cut-off frequency defined by the lower frequency value of the predicted class label. 

\begin{equation}
s_{t,b} (n) = x_t(n) * g_{b} (n)
\end{equation}

\noindent The upper cut-off frequency of filter $g_{b}(n)$ in band $b$ is the upper frequency limit corresponding to the class label. Note that each predicted class label represents a frequency band with upper and lower cut-off frequencies. $g_{b}(n)$ represents a fourth order elliptic filter~\cite{scheirer1998tempo} with minimum transition region. Thus, we can obtain sub-bands of the signal which contains the pitch from the class labels. 

\begin{figure}[t]
				\centering
                \includegraphics[width=\linewidth]{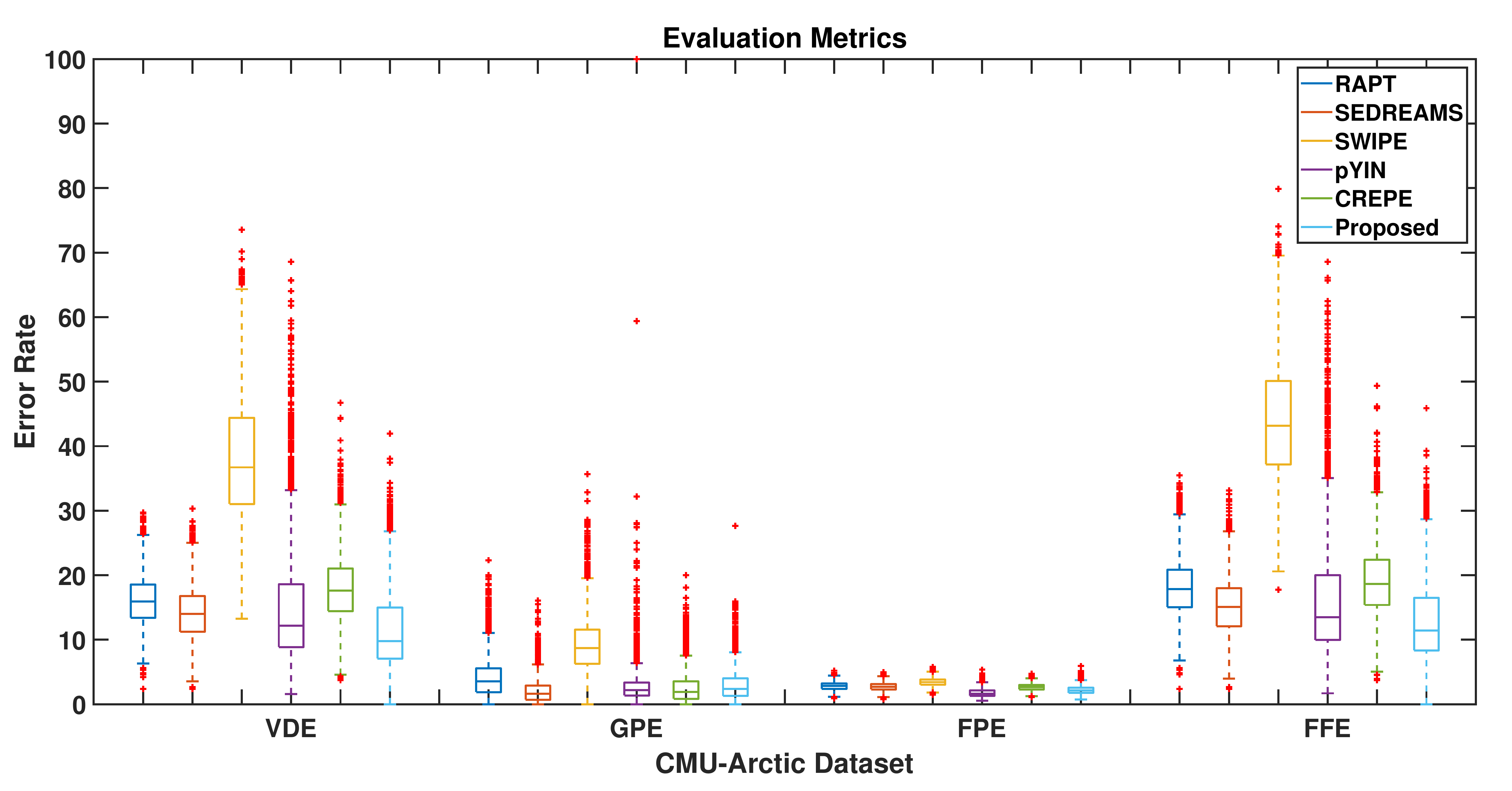}
                \caption{Performance evaluation of $f_0$ extraction methods for CMU-Arctic speech dataset.}
                \label{fig:cmu_eval}
                \vspace{-0.2cm}

\end{figure}

\begin{figure}[t]
				\centering
                \includegraphics[width=\linewidth]{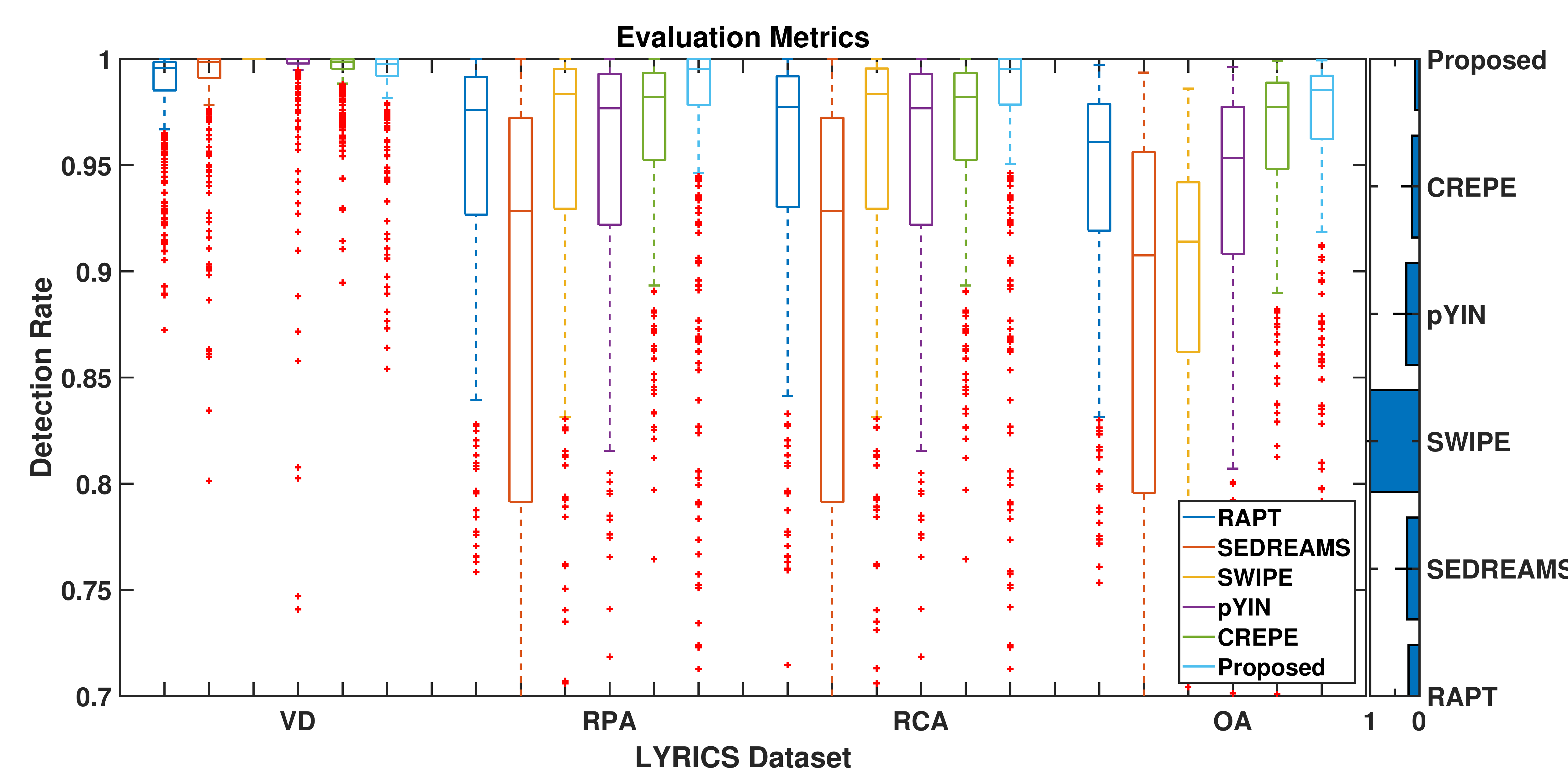}
                \caption{Evaluation of $f_0$ extraction methods for LYRICS dataset. VFA is plotted on right hand side of the figure.}
                \label{fig:lyrics_eval}
                \vspace{-0.2cm}
\end{figure}

\subsubsection{Pitch Extraction from Decoded Bands}
\label{subsubsec:pitch_extract}

The pitch frequency sub-band obtained from the class label is segmented into $50ms$ frame with $80\%$ overlap. The normalized autocorrelation coefficients $r_{t,b}$ of frame $t$ in the sub-band $s_{t,b}$ is computed as

\begin{equation}
r_{t,b}(\tau) = \frac{1}{N}\sum_{j=t+1}^{j=t+N} s_{t,b}(j) s_{t,b}(j+\tau), 0\leq \tau < N
\end{equation}

\noindent Further, to de-emphasize the effect of higher harmonics in the sub-band, we compute autocorrelation on the autocorrelated frames ($r_{t,b}$ denoted as $rr_{t,b}(\tau)$ and further, samples outside the frame size $N$ are considered as zero.

\begin{equation}
rr_{t,b}(\tau) = \frac{1}{N}\sum_{j=t+1}^{j=t+N-\tau} r_{t,b}(j) r_{t,b}(j+\tau), 0\leq \tau < N
\end{equation}

\noindent The candidate fundamental period $t_0$  from $rr_{t,b}(\tau)$ is obtained by   
\begin{equation}
t_0 \approx \left(\underset{\tau \neq 0} {\mathrm{argmax}}~ rr_{t,b}(\tau) \right)
\end{equation}
As the double autocorrelation $rr_{t,b}$ shifts $t_0$ by minimal number of samples, the closest peak to $t_0$ in $r_{t,b}(\tau)$ is identified as $t_0$. The fine pitch error is reduced by performing parabolic interpolation~\cite{de2002yin} over the detected peak in $r_{t,b}(\tau)$. The inverse of $t_0$ is computed to obtain the fundamental frequency.

\begin{figure*}[h]
\centering
\includegraphics[width=\linewidth]{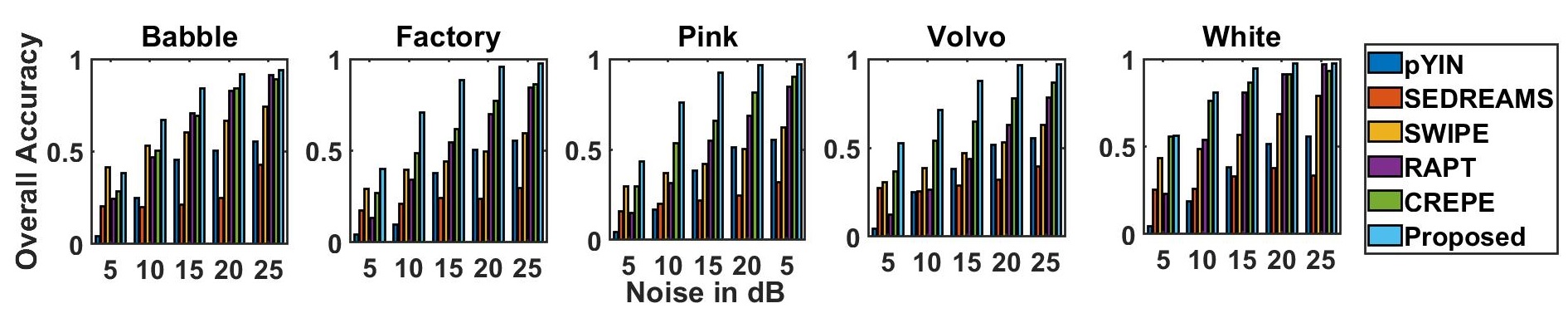}
\vspace{-0.6cm}
\caption{Performance evaluation of the pitch estimation methods under noisy conditions.}
\label{fig_sim}
\vspace{-0.5cm}

\end{figure*}

\section{Datasets}

The openly available datasets: Keele~\cite{Keele} and CMU-Arctic~\cite{CMUArctic} for speech, and LYRICS~\cite{henrich2001study,henrich2005glottal}, MIR-1K~\cite{mir1kdatabase} for singing voice are used to evaluate the proposed method objectively. The emotional/expressive dataset IITKGP-SEHSC~\cite{koolagudi2011iitkgp} is evaluted qualitatively with the spectrograms since the ground truth picth is not available. The Keele dataset (D1) includes ten audio clips with five male and female speakers. The CMU-Arctic dataset (D2) consists of 1131 audio clips for each speaker: BDL (US male), JMK (Canadian male) and SLT (US female). The D1 dataset consists of pitch markers where as the pitch markers for D2 is derived from the simulataneously recorded EGG~\cite{drugman2012detection} available in the dataset. The LYRICS dataset (D3) is a vocal training dataset consists of 437 songs recorded from 13 professional singers. The vocal training includes ornamentations such as crescendos, arpeggios, and glissandos. The MIR-1K dataset (D4) includes 1000 Chinese pop songs rendered by many singers. Three randomly chosen male and female speakers from D1 and 100 randomly chosen songs from D3 are used to train the CNN model to encode pitch frequency band with nominal class labels. The sampled audio signals are split in the proportion 5:3:2 for training, validation and testing respectively. The best performing model is selected after training until the validation accuracy no longer improved for 10 consequitive epochs. Five-fold cross validation is performed to ensure the consistency of the model. During testing, the proposed method is evaluated over partially trained datasets D1 and D3 along with the unseen datasets D2 and D4.

\section{Evaluation and Discussion}

The proposed $hf_0$ is compared with the state-of-the-art (SOTA) singing voice (CREPE~\cite{kim2018crepe} and PYIN~\cite{mauch2014pyin}), and speech (RAPT~\cite{talkin1995robust} and SWIPE~\cite{camacho2011swipe}) $f_0$ extraction methods. The standard evaluation metrics~\cite{speechMetrics,rengaswamy2016robust}: Voicing Decision Error (VDE), Gross Pitch Error (GPE), Fine Pitch Error (FPE) and $f_0$ Frame Error (FFE) are used to evaluate the speech datasets. The evaluation metrics ~\cite{MIREXEvaluation2005}: Voicing Decision (VD), Voicing False Alarm (VFA), Raw Pitch Accuracy (RPA), Raw Chroma Accuracy (RCA) and Overall Accuracy (OA) are used to evaluate the singing voice. Fig.~\ref{fig:cmu_eval} shows the various evaluation metrics on a large scale CMU-Arctic speech dataset for the proposed $hf_0$ and other SOTA $f_0$ extraction methods. From Fig.~\ref{fig:cmu_eval}, we can observe that the VDE of the proposed $hf_0$ is significantly better than the participating SOTA methods. Further, we can observe that the VFA is significantly better than the completely supervised CREPE, which is trained with more than 30 hours of fine pitch labeled data. The GPE and FPE of the $hf_0$ is on par with other methods. Also, we can note that the variance of GPE and FPE of $hf_0$ is very negligible which indicates that $hf_0$ is very stable even though it is not trained with CMU-Arctic dataset. This confirms the adaptability of  $hf_0$ for unseen mode of data. Further, we can observe that $hf_0$ makes significantly less FFE i.e., miss-classifying both voiced/unvoiced frames and making $f_0$ frame errors simultaneously. 

\begin{figure}[h]
			\centering
                \includegraphics[width=\linewidth]{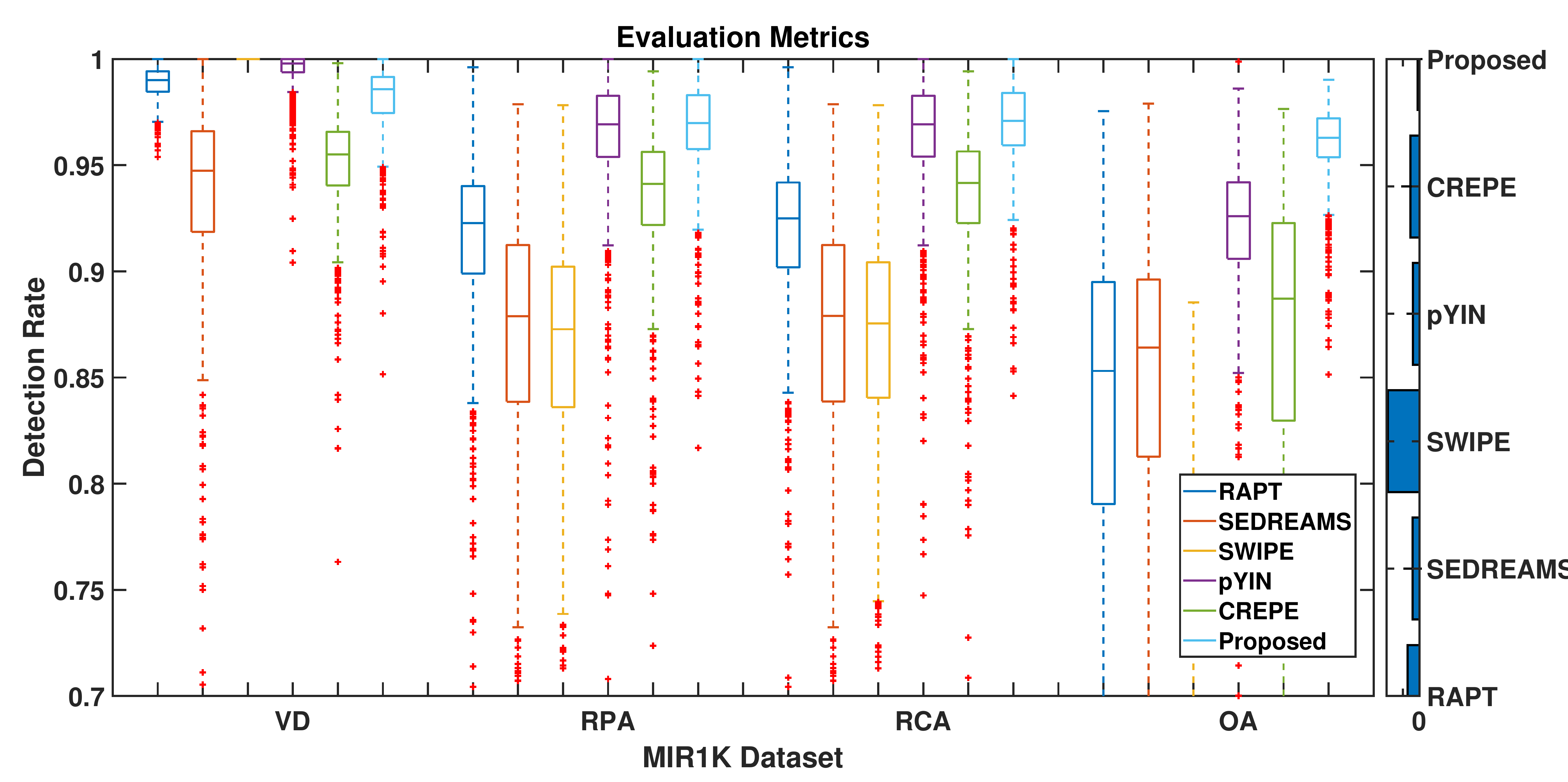}
                \caption{Evaluation of $f_0$ extraction methods for MIR-1K dataset.The VFA is plotted on right hand side of the figure.}
                \vspace{-0.5cm}

                \label{fig:mir_eval}
\end{figure}

The evaluation metrics of $hf_0$ compared with other methods on the LYRICS dataset is shown in Fig.~\ref{fig:lyrics_eval}. LYRICS is one of the complicated dataset with complex ornamentation's such as crescendos, arpeggios, and glissando's with very high singing pitch frequency range. From Fig.~\ref{fig:lyrics_eval}, we can note that $hf_0$ outperforms all other methods in terms of RPA, RCA and OA. We can also observe that $hf_0$ makes significantly less octave errors compare to CREPE and PYIN which are considered as SOTA methods for singing pitch extraction. Further, we can observe that $hf_0$ has significantly high OA (which combines voicing and pitch detection  measures) with negligible variance which confirms that $hf_0$ is very stable to drastic changes in pitch values and the singer pitch range. Similarly, for MIR-1K singing dataset, $hf_0$ is on par with PYIN for RPA and RCA. The OA of the $hf_0$ is significantly better than other methods which indicates that $hf_0$ can be used to extract ground truth pitch for melody and multipitch extraction tasks where ground truth pitch with very high accuracy is crucial. Also, we can note that the VFA of  $hf_0$ is significantly lesser than other methods for both LYRICS and MIR-1K datasets. The qualitative evaluation of $hf_0$ on emotional/expressive data is provided on companion website. The noise evaluation of the $hf_0$ compared with other methods for various noises is shown in Fig.~\ref{fig_sim}. We can observe that $hf_0$ performs significantly better than other methods across all noises and almost for all noise levels.  We can attribute the significant improvement of evaluation measures on various datasets to the CNN model which leans frequency bands instead of pitch class labels results in accommodating wide pitch range of pitch frequency and the unsupervised signal processing method which is invariant to changes in data distribution extracts the pitch based on the fundamental property of the voiced signal production which is invariant to any given periodic signal on the universe.                             

\section{Summary}
In this letters, we proposed a novel hybrid $f_0$ extraction method which exploits the advantages of both traditional signal processing and deep learning approaches. The proposed hybrid $f_0$ extractor is designed to extract the $f_0$ robustly and seamlessly from most common modes of voice production. The qualitative and quantitative evaluation measures showed that the proposed method is significantly better than the state-of-the-art signal processing and deep learning approaches for various modes of voice production.

\ifCLASSOPTIONcaptionsoff
  \newpage
\fi



%

\end{document}